\documentclass[doublecol,figures]{epl2} %for 2 columns style without line numbers
\usepackage{amsmath,amsfonts,amssymb,bm}

\newcommand{\nn}{{\cal N}_x}
\newcommand{\com}{{\cal A}_x}

\title{Rare beneficial mutations cannot halt Muller's ratchet in spatial populations}
\author{Su-Chan Park\inst{1} \and Philipp Klatt\inst{2} \and Joachim Krug\inst{2}}
\shortauthor{S.-C. Park \etal}

\institute{                    
  \inst{1} Department of Physics, The Catholic University of Korea - 43 Jibong-ro, Bucheon 14662, Republic of Korea\\
  \inst{2} Institut f\"ur Biologische Physik, Universit\"at zu K\"oln
  - Z\"ulpicher Str. 77, 50937 K\"oln, Germany
}
\abstract{Muller's ratchet describes the irreversible accumulation of
  deleterious mutations in asexual populations. In well-mixed 
populations the speed of fitness
  decline is exponentially small in the population size, and any
  positive rate of beneficial mutations is sufficient to reverse the
  ratchet in large populations. The behavior is
  fundamentally different in populations with spatial structure, because
  the speed of the ratchet remains nonzero in the infinite size limit
  when the deleterious mutation rate exceeds a critical value. Based
  on the relation between the spatial ratchet and directed
  percolation, we develop a scaling theory incorporating 
both deleterious and beneficial mutations. The theory is
  verified by extensive simulations in one and two dimensions.}

\begin{document}

\maketitle

%% here a revision

%\revision{Insert here the text.
%See fig.~\ref{fig.1}, table~\ref{tab.1} and eq.~(\ref{eq.1}).
%See also~\cite{b.a,b.b}.}

%here a shortcut $\emc$ and again $\emc$

\section{Introduction }
The evolution of asexual populations is driven by novel mutations as
the sole source of genetic diversity. Although the vast majority of
mutations is expected to decrease fitness, these
deleterious mutations often play only a minor role because they
are efficiently purged by natural selection. However, in small
populations deleterious mutations may spread and fix due to stochastic drift. 
When all mutations are deleterious this leads to an irreversible
decline of the fitness of the population that is known as \textit{Muller's
ratchet}. 

The ratchet mechanism was first described verbally by Hermann Muller
in the context of identifying possible evolutionary advantages of genetic
recombination \cite{Muller1964}. Recombination counteracts the
accumulation of deleterious mutations, because the number of mutations
of an individual (its mutational load) can be reduced by
recombining with another individual that carries mutations at different
genetic loci. The standard mathematical formulation of Muller's
ratchet considers an asexual population that is
well-mixed, in the sense that competition between individuals is
implemented only through the constraint of constant population size
$N$. Deleterious mutations occur at rate $U_d$ per individual and
generation, and each mutation decreases the fitness of the individual
by a constant factor that we denote by $e^{-s}$ with $s > 0$ \cite{Felsenstein1974,Haigh1978}. Despite
the simplicity of the model, the computation of the speed of fitness
decline as a function of $N$, $U_d$ and $s$ is a hard problem that
has attracted the attention of population geneticists for more than 40
years \cite{Stephan1993,Gordo2000,Gordo2000a,Rouzine2008,Etheridge2009,Waxman2010,Neher2012,Metzger2013}. 

A key parameter governing the behavior of the ratchet is the deterministic expectation of the 
number of individuals that carry the smallest mutational load, which
is given by  \cite{Haigh1978}
\begin{equation} 
\label{Eq:n0}
n_0 = N e^{-U_d/s}.
\end{equation}
A click of the ratchet occurs when the least loaded class goes extinct and the minimal number of deleterious
mutations carried by any individual increases by one. 
When $n_0 \gg 1$ this event is rare and the fitness declines slowly, whereas for $n_0 \sim
1$ the decline is rapid and continuous. Importantly, for sufficiently
large populations the speed of the ratchet becomes immeasurably small,
irrespective of the values of $U_d$ and $s$. Correspondingly, in large
populations any positive rate $U_b > 0$  of beneficial mutations is sufficient to halt
and reverse the ratchet such that the fitness of the population
increases \cite{Rouzine2008,Yu2010}. 

Here we show that the scenario is fundamentally different in
spatial populations, where the competition between individuals is
limited to their local neighborhood. Spatial models of adaptive
evolution driven by \textit{beneficial} mutations have been developed in
various contexts, and a number of characteristic features have
been identified that differ from the well-mixed setting
\cite{Komarova2006,Gordo2006,Korolev2010,Otwinowski2011,Martens2011,Lavrentovich2013,Lavrentovich2015,Durrett2015,Durrett2016}.
In particular, provided the density of individuals is bounded, the
speed of adaptation remains finite when the habitat size tends to
infinity \cite{Martens2011}. This is in contrast to well-mixed
populations, where the speed of fitness increase diverges
logarithmically with the population size \cite{Rouzine2008,Yu2010,Park2010SK}. 

The effect of spatial structure is even more pronounced for the
accumulation of deleterious mutations. It was shown in
\cite{Otwinowski2014} that the spatial Muller's ratchet in an infinite
habitat displays a sharp phase transition at a critical value $U_d^c$
of the deleterious mutation rate such that the fitness declines at a
finite rate for $U_d > U_d^c$ and remains constant for $U_d < U_d^c$. 
In the following these two dynamical states will be referred to as the 
\textit{moving ratchet} (MR) regime and the 
\textit{halting ratchet} (HR) regime, respectively. 

The existence of a phase transition in the spatial ratchet,
together with the observation that the speed of adaptation is bounded
and vanishes when the beneficial mutation rate tends to zero, suggests 
that a small amount of beneficial mutations will not be able
to halt or reverse the ratchet when $U_d > U_d^c$. 
The purpose of this Letter is to verify and corroborate this
conjecture using extensive simulations in one and two-dimensional
habitats. Based on these simulations and the known relation of the
spatial Muller's ratchet problem to directed percolation and
nonequilibrium wetting \cite{Otwinowski2014}, we develop a scaling 
theory for the (positive
or negative) speed of fitness change as a function of 
the parameters $U_d, U_b$ and $s$. In the next section we introduce
the spatial evolution model and outline its relation to other problems
in nonequilibrium statistical physics. We then present our scaling
theory and the numerical results, and conclude the paper with a summary and a
discussion of some biological implications.

\section{Model}
We first define the spatial model in a general setting and later specify
the rules that are suitable for our purpose. We consider a system with $N$ sites, each of
which is indexed by an integer $x=1,2,\ldots,N$. 
At each site, a single individual is accommodated and the individual
is characterized by its fitness $w_x$.
By $\nn$ we denote the set of indices of nearest neighbors of site $x$.
For later purposes the union of $\nn$ and $\{x\}$ is denoted by $\com$.
Our main interest is in the behavior of a population in the infinite $N$ limit.

We consider a nonoverlapping-generation model with parallel update in
the spirit of Wright and Fisher (WF) \cite{Park2010SK,Wright1931,Fisher1930Book}.
If the fitness at site $x$ is $w_x$ in generation $t$, the fitness
distribution in the next generation $t+1$ is determined in two steps.
\begin{description}
	\item[Selection step:]Fitness at site $x$ is replaced by fitness at site $y$,
where $y$ is chosen among $\com$ according to the probability
\begin{equation}
	S_{x,y}  = w_y \left ( \sum_{z\in \com} w_z \right )^{-1}.
	\label{Eq:sel_prob}
\end{equation}
Needless to say, $y$ can be $x$ itself. The fitness values at all sites are updated simultaneously.
\item[Mutation step:]
After the selection step, the fitness at every site can be modified by mutation.
We denote the mutation probability (density) by $\mu(w'|w)$ which means
that fitness becomes $w'$ if $w$ is the fitness after the selection step.
\end{description}
In this Letter, we limit ourselves to the case 
that fitness takes the form $w_x = e^{s h_x}$, where $s$ is a nonnegative
real number and $h_x$ is an integer,
and mutation can change $h_x$ by an integer value $m$ with probability $U(m)$.
That is, $\mu(w'|w) = \sum_m U(m) \delta\left (w'-we^{sm}\right )$,
where $m$ runs over all integers. 
In the following, $s$ will be called the selection coefficient and
$h_x$ will be referred to as the number of mutations, where
deleterious mutations are counted negatively.
For convenience (and anticipating the connection of this model to surface growth models),
we will also call $h_x$ the height (at $x$).
This mutation scheme is multiplicative in terms of the $w_x$ and
additive in terms of the $h_x$, which implies that epistatic
interactions between mutations are excluded \cite{Jain2008}. 

To be specific, we define $U(m)$ as the convolution of two probabilities $\mu_b(m)$ and
$\mu_d(n)$ with
\begin{align}
	\nonumber
	\mu_d(n) &= \frac{U_d^{-n}}{(-n)!} e^{-U_d},\\ 
	\mu_b(m) &= (1-U_b) \delta_{m,0} + U_b \delta_{m,1},
	\label{Eq:mut}
\end{align}
where $1/k!$ with a negative integer $k$ is interpreted as 0, and
$U_d$ and $U_b$ denote the probabilities of deleterious and beneficial mutations.
The Poisson distribution for deleterious mutations is the typical choice in 
studies of Muller's ratchet~\cite{Haigh1978}, but other forms of
$\mu_d$ will not 
change our conclusions as long as $U_d \ll 1$.

\subsection{Speed of fitness change}
We are mainly interested in the speed of fitness change which is defined as
\begin{align}
	v(t) =\frac{1}{N} \sum_x \left \langle \frac{dh_x}{dt} \right \rangle, \;
	V_h =  \lim_{t\rightarrow\infty}v(t),\;
	V_A = s V_h.
\end{align}
Here the time derivative of $h_x$ should be understood as 
$h_x(t+1) - h_x(t)$ 
and $\langle \ldots \rangle$ stands for an average over all realizations.
Whereas trivially $V_A =0$ if $s=0$, $V_h$ can be nonzero even for $s=0$. We will
therefore focus on the more informative quantity $V_h$ in the
following. For $V_h > 0$ the population adapts whereas for $V_h < 0$
it is subject to fitness decline. 

For the general class of models defined above it can be shown that the speed is 
related to the steady state distribution of the height configurations as \cite{ParkKrugUnpub}
\begin{equation}
	v(t)  = M + \left \langle \sum_{y \in \nn} \left ( h_{{y}} - h_{{x}}\right ) S_{{x},{y}}  \right \rangle,
	\label{Eq:speed}
\end{equation}
where $M$ is the average change of height by mutations per generation per capita, that is, 
$M  \equiv \sum_m m U(m)$, and the average on the right hand side is taken at 
generation $t$.  For the mutation scheme in Eq.~\eqref{Eq:mut}, $M = U_b -U_d$.

This equation has a simple interpretation. 
The contribution to the speed is twofold, by mutation and selection. 
Since mutation and selection are operating separately, the contributions just add up.
Obviously, mutation changes the height by $M$ on average.
Selection can change the height at site $x$ only if a neighbor is chosen as a parent. 
For a given configuration, the increase is $h_{y} - h_{x}$, which
happens with probability $S_{{x},{y}}$. Hence we obtain
Eq.~\eqref{Eq:speed}. 

The spatial structure is determined by $\nn$. 
For the one dimensional system we take $\nn = \{x+1\}$ with periodic
boundary conditions as in Ref.~\cite{Otwinowski2014}.
For two and higher dimensional systems, we distribute the sites on a 
hypercubic lattice with size $N = L^D$ and choose $\nn$ as the
conventional nearest-neighbor neighborhood with periodic
boundary conditions. 
If $\com$ is the set of all indices for any $x$, the model becomes
the well-mixed WF model whose infinite population-size limit has an exact 
solution~\cite{Park2010SK,Park2007K}.
In this case Eq.~\eqref{Eq:speed} is identical to the relation
first obtained by Guess~\cite{Guess1974A} (see also Ref.~\cite{Park2013K}
for the case with recombination). In all the simulations reported below we
will use~\eqref{Eq:speed} to determine the speed, but fitting the mean height by a linear function
gives consistent results.  

\subsection{Relation to other models}
The model without beneficial mutations, that is $U_b=0$, is related to other 
well-studied  models of statistical physics \cite{HHLBook}.   
Consider first the case $U_d = 0$. If $s$ is allowed to take also negative values and $h_x$
is restricted to two consecutive heights  (say, $-1$ and 0) as initial condition, 
the model becomes equivalent to a biased voter model or compact directed percolation~(CDP)~\cite{Lavrentovich2013,Domany1984,Essam1989,Dickman1995}. 
For $s > 0$ ($s < 0$) the system converges to the uniform state $h_x
\equiv 0$ ($h_x \equiv -1$), and the CDP critical  point is located at $s =
s_0 = 0$.
If $U_d$ and $s$ are both positive and we choose the initial condition as $h_x = 0$ for all $x$,
the model becomes a growth model with a nonequilibrium wetting
transition of the kind
first studied in~\cite{Alon1996,Alon1998}. 
The critical behavior of such models can be described by a multilayer extension of
directed percolation (DP) known as unidirectionally coupled 
directed percolation (UCDP)~\cite{Tauber1998,Goldschmidt1999}.
In particular, the depinning of the
surface from the initial level $h_x=0$ that occurs above a critical
value $U_d^c$ is driven by the extinction of a DP process defined
in the $h_x = 0$ layer. 

In the present context this implies that
the deleterious mutation rate $U_d$ mediates a crossover from CDP to  
DP~\cite{Lavrentovich2015}.
According to scaling theory, the critical point is 
expected to be shifted by an amount
$s_c - s_0 \sim U_d^{1/\phi}$ for small $U_d$, 
where $\phi$ is the crossover exponent ~\cite{Janssen2005,Lubeck2006}. 
Since $\phi = 2/D$ for $D<2$, $\phi = 1$ for $D\ge 2$, and $s_0=0$, we get
\begin{align}
	\label{Eq:cross}
	U_d^c \sim 
	\begin{cases} s^{2} & D=1\\
		s & D \ge 2
	\end{cases}
\end{align}
with logarithmic corrections for $D=2$~\cite{Janssen2005}. 
The scaling of $U_d^c$ with $s$ was previously derived and verified 
numerically in \cite{Lavrentovich2013,Otwinowski2014} for $D=1$ and in  
\cite{Lavrentovich2015} for $D=2$.
The fact that $U_d^c \sim s$ for $D>2$ is consistent with the scaling
of $s$ and $U_d$ in the well-mixed population, see
Eq.~\eqref{Eq:n0}. However, in that case there is no phase transition at
any finite value of $U_d$.

\section{Results}
\begin{figure}
	\onefigure[width=0.8\linewidth]{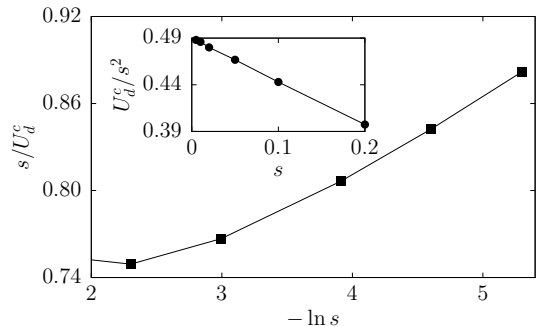}
	\caption{Plot of $s/U_d^c$ vs $-\ln s$ for the  two-dimensional system. 
	Inset: Plot of $U_d^c/s^2$ vs $s$ for the one dimensional system.}
\label{Fig:cross}
\end{figure}
In this section we present and interpret our simulation results.
We first determined the critical point $U_d^c$ for various values of the selection coefficient $s$ 
by exploiting the fact that the density $\rho_0(t)$ of sites with $h_x=0$ decays as
$t^{-\delta}$ at the critical point, where $\delta$ is the critical decay exponent of DP.
The numerical values of $\delta$ are $\approx 0.159~464$~\cite{Jensen1999}
and $0.4510$~\cite{Wang2013} for $D=1$ and $D=2$, respectively. 
We located the critical point by analyzing how $\rho_0(t) t^{\delta}$ behaves with time.
It should veer up (down) if the system is in the HR (MR) regime and saturate to a constant if the system
is at the critical point. 

Using these estimates we verified the validity of the crossover scaling~\eqref{Eq:cross}.
To confirm the logarithmic correction in two dimensions we plot 
$s/U_d^c$ as a function of $-\ln s$ in Fig.~\ref{Fig:cross}. 
In the inset of Fig.~\ref{Fig:cross}, we also plot $U_d^c/s^2$ as a function of
$s$ for the one-dimensional case. 
As $s$ decreases, $U_d^c/s^2$ approaches a constant with a finite slope,
which indicates that the leading term of corrections to scaling is $\sim s$.
That is, $U_d^c \sim s^2 ( c_1 + c_2 s )$ with (nonuniversal) constants $c_1\approx 0.49$ and 
$c_2\approx 0.46$.

\subsection{Scaling theory}
As was mentioned previously, the model with $U_b =0$ shares the (universal) critical
behavior of the nonequilibrium wetting models of Refs.~\cite{Alon1996,Alon1998}.
When $U_d \neq U_d^c$, the system has a single time scale
$|U_d - U_d^c|^{-\nu_\|}$, where $\nu_\|$ is the correlation time
exponent of DP. Numerical values of $\nu_\|$ are 1.733~847~\cite{Jensen1999} and 1.287~\cite{Wang2013} for
one and two dimensions, respectively. If $U_d > U_d^c$, the ratchet is moving and
the speed is proportional to the inverse of the characteristic time scale,
because this is the time scale on which the currently least loaded type
goes extinct. Accordingly, the speed in the MR regime
is $V_h \sim (U_d - U_d^c)^{\nu_\|}$~\cite{Otwinowski2014}.

\begin{figure}
	\onefigure[width=0.8\linewidth]{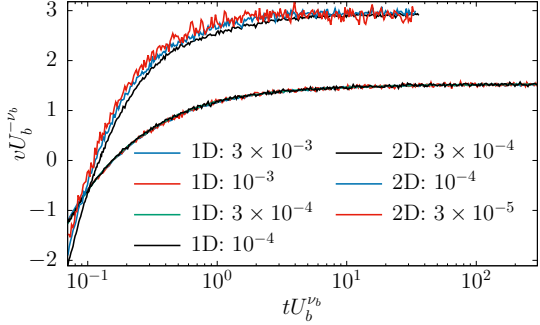}
	\caption{\label{Fig:mupara} Semilogarithmic plots of $v U_b^{-\nu_b}$ vs $t U_b^{\nu_b}$ 
	at the critical point with $\nu_b = 0.76$ and 0.81 for the
	one-dimensional (bottom) and two-dimensional (top) systems,
        respectively. The selection coefficient is $s=2.5$ and the
        corresponding deleterious mutation rates are $U_d^c \approx 0.304725$
        ($D=1$) and $U_d^c \approx 1.215433$ ($D=2$). The values for
        the beneficial mutation rate $U_b$ are displayed in the figure.  
	}
\end{figure}
Based on these considerations we can make a scaling ansatz
\begin{align}
	v(t) = |\Delta_d|^{\nu_\|} G_0\left ( \Delta_dt^{1/\nu_\|}  \right ), 
\end{align}
where $\Delta_d \equiv U_d-U_d^c$, and $G_0$ is a scaling
function with the (anticipated) asymptotic behavior 
\begin{align}
	G_0(x) \sim \begin{cases}
		|x|^{-\nu_\|}, & x\rightarrow 0,\\
		\text{(negative) const}, &x \rightarrow \infty,\\
		0, & x\rightarrow -\infty.
	\end{cases}
\end{align}
Note that this predicts that the speed decays as $t^{-1}$ 
at the critical point, which corresponds to the logarithmic growth of the mean height~\cite{Hinrichsen2003}.

For nonzero $U_b$, the scaling ansatz needs to be extended to incorporate the effect of beneficial
mutations. Assuming that $U_b$ affects the speed in a power-law fashion, we can write
\begin{align}
	v(t) = t^{-1} F \left ( \Delta_d t^{1/\nu_\|} , t U_b^{\nu_b} \right ),
\end{align}
where $\nu_b$ is a new exponent and the relation between $F$ and $G_0$ is $F(x,0) = G_0(x) |x|^{\nu_\|}$.
At the critical point ($\Delta_d = 0$), we rewrite $v(t)$ as
\begin{align}
	v(t) = t^{-1} F \left ( 0, t U_b^{\nu_b} \right )
	=U_b^{\nu_b} G\left ( t U_b^{\nu_b} \right ),
\end{align}
where $G(x) = F(0,x)/x$. The scaling ansatz suggests that a data collapse is expected when
$v U_b^{-\nu_b}$ is plotted against $t U_b^{\nu_b}$ for various $U_b$'s.
In Fig.~\ref{Fig:mupara}, we indeed observe a data collapse when we set $\nu_b = 0.76$ ($\nu_b = 0.81$) for
the one-dimensional (two-dimensional) system. 
By observing where the scaling collapse becomes worse, 
we conclude
that $\nu_b = 0.76 \pm 0.03$ in one dimension
and $\nu_b = 0.81 \pm 0.03$ in two dimensions,
and the behavior of the speed $V_h$ at the critical point for nonzero $U_b$ is
$V_h \sim U_b^{\nu_b}$. 
The seeming deviation of the collapse of the 2D system in comparison to the 1D system is due to
strong corrections to scaling in the two-dimensional case.

At the critical point, $t_b \equiv U_b^{-\nu_b}$ is the unique characteristic time scale of the system.
If the system is not at the critical point, there is another time scale 
$t_c \equiv |\Delta_d|^{-\nu_\|}$.
Since the speed $V_h$ is inversely proportional to the characteristic time scale (as long as
it is nonzero), the behavior of $V_h$ will be 
determined by the smaller of the two time scales. 
If $t_b \ll t_c$, $V_h$ is dominated by the behavior at the critical point, whereas
if $t_b \gg t_c$ it is dominated by the off-critical behavior for $U_b = 0$. 
In this sense, there is a crossover from the
DP type behavior to a behavior dominated by beneficial mutations at
$t_b \simeq t_c$, or $U_b^{1/\varphi} \simeq |\Delta_d|$, where 
$\varphi = \nu_\|/\nu_b$ 
is a crossover exponent.
In one (two) dimension, we get $\varphi \approx 2.28$ (1.59).
From these considerations, the asymptotic speed $V_h$ is expected to take the form
\begin{align}
	V_h = c_s^{-1} U_b^{\nu_b} H\left ( a_s \Delta_d  U_b^{-1/\varphi} \right ),
	\label{Eq:speed_ansatz}
\end{align}
where $H$ is a universal function with geometric constants $c_s$ and
$a_s$, which are determined by requiring $H(1) =0$ and $H(0)=1$.

\begin{figure}
	\onefigure[width=0.8\linewidth]{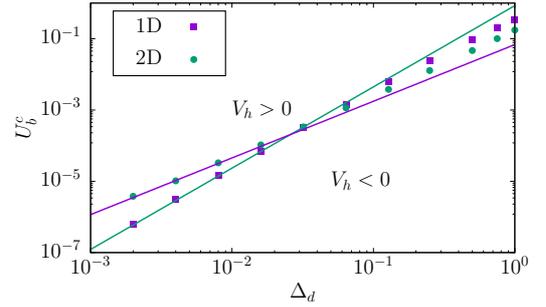}
	\caption{\label{Fig:PB} Plots of $U_b^c$ vs $\Delta_d$ for one- (square) and two-dimensional (circle) systems
	for $s=2.5$. The straight lines show Eq.~\eqref{Eq:Ubc} 
	with the numerical values for $\varphi$ mentioned in the text.}
\end{figure}
The scaling ansatz~\eqref{Eq:speed_ansatz} suggests another way of finding $\varphi$ by studying how the solution $U_b^c$ of 
$V_h(U_b)=0$ behaves as $\Delta_d$ varies. 
Since $U_b^c$ quantifies the beneficial mutation rate that is required to reverse the ratchet,
we will refer to it as the turning point. From the scaling ansatz, we expect
\begin{equation}
\label{Eq:Ubc}
	U_b^c \approx  (a_s \Delta_d)^{\varphi}.
\end{equation}
In Fig.~\ref{Fig:PB}, we depict $U_b^c$ as a function of $\Delta_d$ for $s=2.5$. The power-law behavior 
for small $\Delta_d$ is consistent with the numerical estimate of $\varphi$ mentioned above.

\begin{figure}
	\onefigure[width=0.8\linewidth]{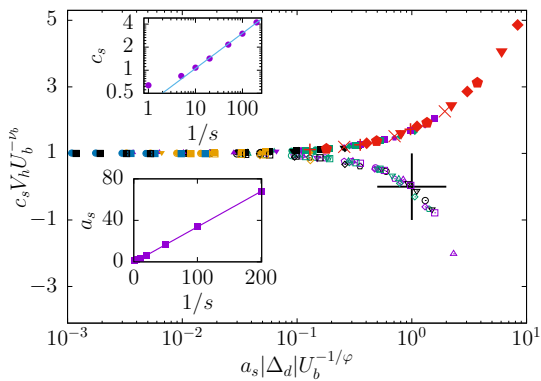}
	\caption{Scaling collapse plot of $c_s V_h U_b^{-\nu_b}$ vs $a_s |\Delta_d| U_b^{-1/\varphi}$ 
	for the one dimensional model for $s=2.5$ (square),
	$1$ (circle), $0.2$ (triangle), 0.1 (inverted triangle), 0.05 (diamond), 0.02 (pentagon), 0.01 ($\times$),
	and 0.005 ($+$) on a semilogarithmic scale. 
	Solid (empty) symbols are results in the HR (MR) regime.
	The large cross formed by a vertical and a horizontal line
        segment marks the position of the turning point, which is
        (1,0) in the rescaled variables.
	Large symbols (in red color) indicate simulation results without deleterious mutations.
	Inset: (bottom) Plot of $a_s$ vs $1/s$. (top) Double logarithmic plot of $c_s$ vs $1/s$. The straight line
	is the result of a two-parameter fit $c_s \approx C s^{-\gamma}$ with $\gamma \approx 0.46$ and $C=0.35$.
	}
\label{Fig:col1}
\end{figure}

\begin{figure}
	\onefigure[width=0.8\linewidth]{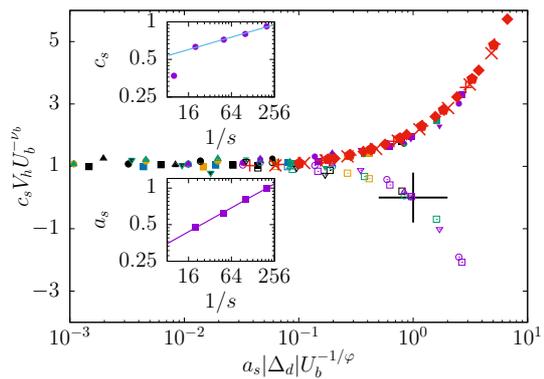}
	\caption{Scaling collapse plot of $c_s V_h U_b^{-\nu_b}$ vs $a_s |\Delta_d| U_b^{-1/\varphi}$ 
	for the two-dimensional model on a semilogarithmic scale.
	We use the same symbols as in Fig.~\ref{Fig:col1} for different values of $s$.
	Inset: Double logarithmic plots of (bottom) $a_s$ and (top) $c_s$ vs $1/s$.  The straight lines
	are results of power-law fittings with slopes $\alpha = 0.33$
        and $\gamma = 0.16$ for $a_s$ and $c_s$, respectively.
	}
\label{Fig:col2}
\end{figure}

In Figs.~\ref{Fig:col1} and \ref{Fig:col2}, we show the scaling collapse predicted by~\eqref{Eq:speed_ansatz} for
one- and two-dimensional systems, respectively. 
The number of individuals is $N=2^{18}$ ($2^{20}$) for the one-dimensional (two-dimensional) simulations.
The number of independent runs for each data point is in the range $400 - 1000$ for the one-dimensional model
and $5000 - 7000$ for the two dimensional model. 
The initial condition is always $h_x = 0$ for all $x$.
By adjusting the geometric constants $a_s$ and $c_s$ for different $s$, all data 
are indeed collapsed into a single curve for various values of $s$.
The curves possess two branches corresponding to the MR and 
HR regimes. Along the lower MR branch the speed changes sign at the turning point $U_b^c$. 

We also analyzed how $a_s$ and $c_s$ behave for small $s$.
As shown in the insets of Figs.~\ref{Fig:col1} and \ref{Fig:col2}, 
we found power-law behaviors $a_s \sim s^{-\alpha}$  and $c_s \sim s^{-\gamma}$ 
with $\alpha \approx 1$ ($0.33$) and $\gamma \approx 0.46$ (0.16) for $D=1$ ($D=2$).
Including also the prefactor in the relation for $a_s$ we conclude that the turning point in the 
one-dimensional system is located approximately at 
$U_b^c \approx (0.34 \Delta_d/s)^{2.28}$. 
Due to the substantial logarithmic corrections, obtaining an accurate
approximation formula for $U_b^c$ in two dimensions would require
more extensive simulations.

\subsection{Relation to DP exponents} 
The numerical estimates of $\nu_b$ presented above are very close to the value of
$1/(1+\eta)$, where $\eta$ is the initial slip exponent of the DP universality class,
with $\eta \approx 0.313~686$  [$1/(1+\eta) \approx 0.7612$]
in one dimension~\cite{Jensen1999} and $\eta \approx 0.2307$ [$1/(1+\eta) = 0.813$] in two dimensions~\cite{Wang2013}.
Here we present an argument in favor of this relation.

For this we need to recall the relation of the model with $U_b = 0$ to UCDP \cite{Tauber1998,Goldschmidt1999}. 
The connection is clearly seen by defining that site $x$ is occupied by species $A_m$ if $h_x \ge -m$ ($m\ge 0$).
Each species undergoes its own DP process. In order
to maintain a well-defined height function it is assumed that a site occupied by species $m$ is also 
occupied by all species $l > m$. In our setting the coupling between the species is not strictly unidirectional,
because the probability that $A_m$ increases by selection is affected by species with $l > m$. 
However, this feedback, which is present also in some nonequilibrium wetting models, was argued to be  
irrelevant in the renormalization group sense~\cite{Goldschmidt1999}. 

By contrast, beneficial mutations induce a direct reverse coupling by which particles of species $m$ induce
the creation of particles of species $l < m$. 
We start from the observation that the critical behavior of the species at the 
lowest level is identical to DP.
Let us start the process at height $h_x = 0$ and place the system at the critical point $U_d^c$ with $0 < U_b \ll 1$.
At generation $t=1$, the density of sites with $h_x = 1$ is then roughly $U_b$. 
In terms of the UCDP, the new species $A_{-1}$ now becomes the lowest level species which
should perform the defect dynamics of DP at the critical point. 
It follows that the density $\rho_{-1}$ of species $A_{-1}$ at generation $t$ is
$\rho_{-1}(t) \sim U_b t^\eta$. 
Notice that $\eta$ already incorporates the effect of deleterious mutations, which correspond to the death of $A_{-1}$ particles.
Since $A_{-1}$ can change into $A_{-2}$ by
another beneficial mutation which occurs with probability $U_b$, 
the expected density of species $A_{-2}$ up to generation $t$ is
$\rho_{-2}(t) \sim U_b^2 t^{1+\eta}$.
If we define $t_b$ as the time when $\rho_{-2}(t_b) = U_b$, which satisfies 
$t_b \sim U_b^{-1/(1+\eta)}$, species $A_{-2}$ at generation $t_b$ takes 
the place of $A_{-1}$ at generation $t=1$.
Hence, we conclude that the mean height of the whole system increases
by one at generation $t_b$, which gives $V_h \sim U_b^{1/(1+\eta)}$ at the critical point
and therefore $\nu_b  = 1/(1+\eta)$.
Since we neglected the effect of increasing $A_{-2}$ by its own dynamics as well as the dynamics of
higher level species, this argument is not exact.  
Nevertheless our numerical estimates lend strong support to the proposed scaling relation.

\subsection{Asymptotics of the scaling function}

To complete the scaling theory, it remains to determine the asymptotic behaviors of the scaling function $H(x)$ in 
Eq.~\eqref{Eq:speed_ansatz}. This is straightforward in the MR regime ($x > 0$). 
Because $V_h \sim \Delta_d^{\nu_\|}$ for $U_b = 0$ and $\Delta_d >0$,
it is expected that $H(x) \sim x^{\nu_\|}$ for large $x$. As a consequence, 
$V_h$ becomes independent of $U_b$ in this regime.

\begin{figure}
	\onefigure[width=0.8\linewidth]{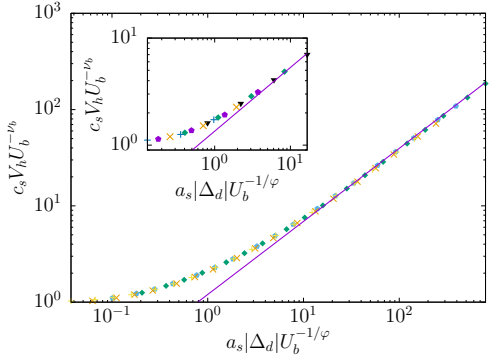}
	 \caption{Double logarithmic plot of $c_s V_h U_b^{-\nu_b}$ vs $a_s |\Delta_d| U_b^{-1/\varphi}$ 
	 for the two dimensional model at $U_d = 0$ ($\Delta_d =-U_d^c$). 
	 The slope $\chi$ of the straight line takes the value predicted by Eq.~\eqref{Eq:chi}.
	The inset shows the same plot for the one dimensional case.
	}
\label{Fig:asymH}
\end{figure}

The behavior for $x \to -\infty$ can be inferred from known results for the case of adaptation in the presence
of purely beneficial mutations \cite{Martens2011}.
Since selection should dominate when $U_d \ll U_b$, it is plausible to assume that the speed is continuous at $U_d = 0$ for
positive $s$ and $U_b$. This suggests that  
the speed for the model without deleterious mutations should also conform to 
the scaling ansatz Eq.~\eqref{Eq:speed_ansatz}, as long as $s$, $U_b$, and $U_d^c$ are small. 
Figures~\ref{Fig:col1} and \ref{Fig:col2} include simulation results for $U_d = 0$ (with bigger symbol size) and 
indeed confirm this anticipation. 
Assuming therefore that the case $U_d = 0$ ($\Delta_d = - U_d^c$) is representative of the 
$\Delta_d < 0$ (HR) regime, we conclude that the scaling $V_h\sim U_b^{1/(1+D)}$ 
known from the model without deleterious mutations~\cite{Martens2011} should be recovered when
$|\Delta_d| U_b^{-1/\varphi}$ is large. 
Thus, we expect for large $-x$ that $H(x) \sim |x|^\chi$ with 
\begin{equation}
\label{Eq:chi}
\chi = \nu_\| \left( 1 - \frac{1}{\nu_b (D+1)} \right) = \frac{\nu_\|(D-\eta)}{D+1},
\end{equation}
where the scaling relation between $\nu_b$ and $\eta$ has been used in the second step.
Accordingly, we get
\begin{align}
	H(x) \sim
	\begin{cases} 
		|x|^\chi & x \rightarrow -\infty\\
		\text{const} & x\rightarrow 0\\
		x^{\nu_\|} & x\rightarrow \infty.
	\end{cases}
	\label{Eq:asymH}
\end{align}
The behavior for $x \to -\infty$ is confirmed by the simulation results in Fig.~\ref{Fig:asymH}. Note that the applicability of the scaling form 
\eqref{Eq:speed_ansatz} to the model without deleterious mutations
implies the existence of a second, previously
unnoticed scaling regime for adapting spatial populations that appears when $U_b$ is large in the sense of 
$a_s U_d^c U_b^{-1/\varphi} \sim s^{\phi-\alpha} U_b^{-1/\varphi} \ll
1$ \cite{ParkKrugUnpub}. The onset of this regime is discernible in
the departure of the data from the straight line in  Fig.~\ref{Fig:asymH}. 

\section{Summary and conclusions}
In this paper we have studied the speed of adaptation or fitness decline of a population with spatial
structure when both beneficial and deleterious mutations are present.
We do not take epistasis into account, not only because the absence of epistasis 
is a prerequisite for the existence of a constant asymptotic speed of fitness change, but also because
epistasis provides an alternative and independent mechanism by which Muller's ratchet 
can be halted \cite{Jain2008}. 
Within this setting, we found a general formula Eq.~\eqref{Eq:speed} 
which allowed us to accurately estimate the speed from simulations without invoking any extrapolation.

When both types of mutations are present the speed turns out to take
a rather complicated form. In particular, it does not reduce to the sum of the two speeds caused by beneficial or deleterious
mutations acting in isolation, as would be the case in small populations where mutations spread and fix independently
\cite{Klatt2017Book}. 
We developed a scaling theory based on the critical behavior of directed 
percolation and its unidirectionally coupled multi-species extension.
The predicted scaling relations are confirmed numerically for $D=1$ and $D=2$, but the theory
should apply in any dimension. 

The dramatic enhancement of the efficacy of Muller's ratchet in spatial habitats that we describe 
is a consequence of the general weakening of natural selection when competition is local rather than global. 
Experiments aimed at verifying the ratchet 
mechanism have mostly been conducted in well-mixed populations 
without spatial structure. The first experiments used RNA viruses, 
which are distinguished by their large deleterious mutation rates $U_d \sim {\cal{O}}(1)$ \cite{Chao1990}. 
Since the corresponding selection coefficients are relatively small \cite{Elena1999}, these 
systems would be predicted to operate deep in the MR regime. 

Deleterious 
mutation rates in bacteria are much lower. For example, the estimates 
$U_d = 5 \cdot 10^{-3}$ and $s = 0.03$ were obtained for a mutator strain 
of \textit{Escherichia coli} \cite{Trindade2010}. 
Although our simplified model cannot be expected to be quantitatively
applicable to this specific microbial system, it is nevertheless instructive to compare these values 
to the critical deleterious mutation rate $U_d^c$ obtained from our simulations. 
This comparison would place the \textit{E. coli} strain 
in the MR regime for 1D populations and in the HR regime for 2D populations. Moreover, using
Eq.~\eqref{Eq:Ubc} we find that a beneficial mutation rate of $U_b \approx 1.2 \cdot 10^{-3} \approx 0.25 \cdot U_d$ would be required
to reverse the ratchet in one-dimensional populations. Whereas two- and three-dimensional habitats are naturally realized
in bacterial colonies, effectively one-dimensional
geometries appear, e.g., at the edge of expanding microbial colonies  
\cite{Lavrentovich2013,Lavrentovich2015,Lavrentovich2016,Bosshard2017}. 

We are therefore confident that an experimental
test of our predictions is principally feasible. In fact, two recent experiments have provided direct evidence for the
enhanced effect of deleterious mutations in spatial habitats \cite{Lavrentovich2016,Bosshard2017}, and it has been suggested
that the exceptionally low mutation rates in bacteria may have evolved as a consequence of biofilm formation, which requires 
a more stringent control of deleterious mutations \cite{Gralka2016}. Taken together, these results indicate an important
role for deleterious mutations in spatial habitats which should be explored further in experimental and theoretical work. 

\acknowledgments
S-CP acknowledges the support by the Basic Science Research Program through the
National Research Foundation of Korea~(NRF) funded by the Ministry of
Science and ICT~(Grant No. 2017R1D1A1B03034878) and by the Catholic University of Korea,
research fund 2018. JK was supported by DFG within SPP 1590 and CRC 1310.
We furthermore thank the Regional Computing Center of the
University of Cologne (RRZK) for providing computing time on the DFG-funded High
Performance Computing (HPC) system CHEOPS.

\bibliographystyle{./eplbib.bst}
\bibliography{Krug}

\end{document}